\documentstyle{article}
\title{POINTLESS SPACES IN GENERAL RELATIVITY}

\author{G.N. Parfionov, R.R. Zapatrin,\\
Friedmann Laboratory for Theoretical Physics,\\
SPb UEF, Griboyedova 30/32,\\
191023, St-Petersburg, Russia}
\date{}
\begin{document}

\maketitle

\begin{abstract}
The new approach to quantize the gravity based on
the notion of differential algebra is suggested. It is shown that
the differential geometry of this object can not be described in
terms of points. The spatialization procedure giving rise to points
by loosing a part  of the  entire  structure  is  discussed.  The
counterpart   of   the traditional objects of differential geometry
are studied.
\end{abstract}

\noindent {\bf FOREWORD}

In general relativity an event in spacetime is idealized  to  a
point of a four-dimensional manifold. Such idealization is
adequate within  classical  physics,   but   is   unsatisfactory
from   the operationalistic point of view. In quantum theory the
influence of a measuring  apparatus  on  the  object  being
observed  can  not  in principle be removed. We could expect  the
metric  of  a  quantized theory to be subject to fluctuations,
whereas the  primary  tool  to separate individual events is  just
the  metric.  Thus  a  sort  of {\it smearing procedure}  for
events is to be imposed into the theory.

\medskip
An essential step in this direction was the idea to  build  the
differential geometry in terms of abstract algebras.  Geroch
(1972) proposed to generalize the notion of algebra of smooth
functions  on a manifold to that of {\it Einstein algebra} whose
elements  are  not  yet functions. This  generalization  was
successful  since  the  entire content of general relativity can be
reformulated in such a way that the underlying spacetime manifold
is used only once: to  define  the collection of smooth functions.

\medskip
Although, since the {\it commutative} case is considered, the absence
of points is, roughly speaking, an illusion. As a matter of fact,
a {\it commutative} algebra can always be represented  by
functions  on  an underlying space. Such representation is, for
instance, the Gel'fand construction (for normed algebra)  which  is
the  special  case  of representation of commutative algebra on its
spectrum.  So,  in  the case of the commutative algebra points
implicitly exist. We consider it in more detail in section 2.

\medskip
The goal of this paper is to essentially remove points from the
theory. Metaphorically speaking, instead of smearing {\it out}  of
events we smear them {\it off}.  This  happens  automatically  when
we  pass  to non-commutative  Einstein  algebras.  Whereas  the
reproduction  of geometrical constructions causes a  number  of
purely  mathematical obstacles. The analysis of these problems is
concluded by an example of a finite dimensional non-commutative
Einstein  algebra  (section 8).

\medskip
\section{POINT-FREE APPROACH TO DIFFERENTIAL GEOMETRY}

\medskip
\medskip
The emphasis of this section is made on  the  observation  that
the standard coordinate-free approach to  differential  geometry  of
smooth manifolds can be thought of as (or converted to) point-free.

\medskip
The basis of the differential geometry is the notion of  {\it vector
field}. It is known any vector field $v$ can  be  associated  with  the
differential operator in the algebra ${\cal A}$ of smooth  functions  on
the manifold acting as the derivation  along  this  vector  field.
This operator $v$ is linear, and its main feature is the {\it
Leibniz rule}:

$$
v(ab)=v(a)b+av(b)
\eqno{(1.1)}
$$

\noindent It is  known  that  linear  operators  in ${\cal A}$
satisfying  (1.1)  are exhausted by that induced by actions of
vector fields. That  is  why the difference is  not  drawn  between
such  operators  and  vector fields: this is  the  essence  of  the
coordinate-free  account  of differential geometry. As a matter of
fact, coordinates appear  only once: to specify the algebra ${\cal A}$
of smooth functions, since the notion of smoothness is referred
to local  maps.  The  forthcoming  notions such as connection,
torsion, curvature  and  others  need  no  local coordinates in
their definition.

\medskip
We  emphasize  that  at  the  mere  level  of  definitions  the
principle notions of differential geometry require  no  coordinates,
 nor even {\it points}: the fact that ${\cal A}$ is the algebra of
functions on  a set is never used. Thus the global geometry {\it
per se} does not confine us by set-theoretical concept of space.

\medskip
\section{TOWARDS NONCOMMUTATIVITY}

\medskip
\medskip
In this section we shall analyze the obstacles arising  in  the
non-commutative generalization  of  the  algebraic  construction  of
differential geometry.

\medskip
\noindent {\bf Basic algebra.} The first question is  why  are  we  going  to
fetch
non-commutativity to geometry? The rough answer is  that  we  follow
the tradition of quantization. An amount of non-commutativity in the
geometry  itself  is  needed  to  quantize  it.  This  produces  the
following problem: the lack  of  points  in  this  quantum  geometry
requires a "spatialization" procedure to be imposed into the general
scheme to describe the {\it observable} entities.

\medskip
We  shall  start  with  an  associative  and,   in   generally,
non-commutative algebra ${\cal A}$ over real numbers which will play the
role analogous to that of algebra of smooth functions. It will be
called the {\it basic algebra} of the model.

\medskip
\noindent {\bf Spatialization procedure}. Let us try to extract
the  geometry  from the basic algebra ${\cal A}$ on its coarsest
level, that is,  set-theoretical one. As it is usually done, we
must consider the elements  of ${\cal A}$  as functions defined on
a certain set $M$, and perhaps, taking the values in a
non-commutative domain {\it R}. That is, the representation of
${\cal A}$  by means of homomorphism $\hat{ }$ is introduced:

$$
a\mapsto\hat{a}
$$

\noindent where $\hat{a}$ is a function $M\rightarrow R$. Thus each
point $m\in M$  is  associated  with the two-sided ideal
$I(m)\subseteq {\cal A}$:

$$
I(m) = \{a\in {\cal A}\mid\hat{a}(m)=0\}
$$

Now we see that the resources of spatialization
are bounded  by the number  of  two-sided  ideals  in ${\cal A}$.
Whereas,  if ${\cal A}$  contains two-sided ideals, it can be, as a
rule,  decomposed  into  mutually commuting components. So, each
point can be associated with at least a simple component of the
decomposition of ${\cal A}$. The conclusion is that {\it
spatialization   and   non-commutativity   are   in    some
sense complementary}: commutation relations can not be described
in  terms of points.

\medskip
When the basic algebra ${\cal A}$  is  commutative  and  satisfies
some additional requirements   (is   Banach   algebra),  the
proposed construction is just the Gel'fand representation endowing
the set $M$ by a natural topology. So, the commutative case makes
it possible to store the {\it topological} space $M$ so that ${\cal
A}$ is represented by continuous functions on {\it M}. However, the
Gel'fand construction  does  not  yield the differential structure
for {\it M}.

\medskip
\noindent {\bf Differential structure.} The lack of points is  not
an  obstacle  to introduce differential structure with all its
attributes. As in  the commutative case, it is introduced in terms
of the  collection  ${\rm Der}{\cal A}$ of derivations of the basic
algebra ${\cal A}$. Recall that a {\it derivation} of ${\cal A}$ is
the linear mapping $v:{\cal A}\rightarrow {\cal A}$ enjoying the
Leibniz rule  (1.1). ${\rm Der}{\cal A}$ is the Lie algebra over
the field ${\cal R}$ of real numbers with respect to the
commutation:

$$
[u,v]a = u(va)-v(ua)
$$

\noindent {\bf Scalars.} In the commutative case we can multiply
a  vector  by  any element of the basic  algebra ${\cal A}$.  In
general,  an  element $v\in {\rm Der}{\cal A}$ multiplied by an element
$a\in {\cal A}$  does  not  enjoy  the  Leibniz  rule.  However, to
define such object as, say,  connection,  multiplicators are
necessary: they play the role of {\it scalars}. So, we have  to
clear out which elements of ${\cal A}$ can serve  as
multiplicators  for  vectors.  Evidently, each element of the
center $Z({\cal A})$ of the  algebra ${\cal A}$  suits for this
purpose: for each $z\in Z({\cal A})$, $v\in {\rm Der}{\cal A}$, $a,b\in
{\cal A}$ the Leibniz formula holds:

$$
(ab) = z(v(a)b+av(b))
= (zv)(a)b+a(zv)(b)
$$

In  the  sequel  we  shall  be
confined  by  this   class   of multiplicators, that is, ${\rm Der}{\cal
A}$ will be considered as $Z({\cal A})$-module. So, $Z({\cal A})$
will be set up as the set of scalars:

$$ S = Z({\cal A}) $$

Note that the set $S$ of multiplicators for $V$ may be  essentially
broader than $Z(A)$, and even non-commutative, whereas  we  shall  not
tackle this problem in our paper since such level of  generality  is
not needed for the account of the proposed model.

\medskip
\section{DIFFERENTIAL ALGEBRAS}

\medskip
\medskip
We introduce the notion  of  {\it differential  algebra}  as  a
pair $({\cal A},V)$, $V\subseteq {\rm Der}{\cal A}$. The reasonable
restrictions on the choice  of $V$  are analyzed in this section.

\medskip
There are natural classical examples with $V\neq {\rm Der}{\cal A}$. The
elements of ${\rm Der}{\cal A}$ are the direct generalization of  vector
fields on  smooth manifolds. Sometimes, even in the classical
(commutative) situation, not all vector fields are considered. For
example, if the algebra of smooth vector fields on a Lie group is
studied, it is natural to  be confined by left invariant  ones.
Another  example  is  yielded  by dynamical systems associated with
the subalgebras of ${\rm Der}{\cal A}$ with  one generator. In classical
mechanics, to fix  up  a  subalgebra $V\subseteq {\rm Der}{\cal A}$
means to define the virtual shifts of the system.

\medskip
\noindent {\bf Constants.} Now let a subset $V\subseteq {\rm Der}{\cal
A}$ is  set  up  whose  elements  are thought  of  as  "virtual
infinitesimal   shifts".   The   question immediately arises which
elements of ${\cal A}$ are invariant with respect to all these
shifts.  Call  such  elements  {\it constants}.  The  set ${\cal
K}$  of constants

$$
{\cal K} = V^{c} = \{k\in {\cal A}\mid \forall v\in V \quad vk=0\}
\eqno{(3.1)}
$$

\noindent is always the subalgebra of ${\cal A}$ (proof is
straightforward).  Clearly, $v(ka)=k\cdot va$ for each $v\in V$,
$a\in {\cal A}$, $k\in {\cal K}$. The counterpart of ${\cal K}$ in
classical mechanics is the algebra of integrals of dynamical
system.

\medskip We emphasize that the set of constants
depends substantially on the choice of $V$. It follows from
(3.1)  that ${\cal K}=V^{c}$  shrinks  when $V$ broadens. In
particular,  when $V={\rm Der}{\cal A}$  we  call  the  elements  of
$C_{\cal A}=({\rm Der}{\cal A})^{c}$ {\it basic  constants}. $C_{\cal A}$
lies in  all  other  algebras  of constants: $C_{\cal A}\subseteq
V^{c}$.  When ${\cal A}$  is  not  commutative $C_{\cal A}$  is
nevertheless commutative and, moreover, is contained in the  center
$Z({\cal A})$  of ${\cal A}$.  Note that $C_{\cal A}$ always
contains the elements of the form $\lambda \circ 1$, $\lambda \in
{\cal R}$.  For any $u,v\in {\rm Der}{\cal A}$, $c\in C_{\cal A}$, $a\in
{\cal A}$

$$
[cu,v]a = cuv(a)-v(cu(a))= cuv(a)-vc\cdot u(a)-cvu(a) = c[u,v]a
$$

\noindent therefore

$$
[cu,v] = c[u,v] = [u,cv] \eqno{(3.2)}
$$

\noindent (the second
equality is proved likewise). Hence, ${\rm Der}{\cal A}$ may be thought
of as the Lie algebra over $C_{\cal A}$. The following example
shows that $C_{\cal A}$ may be broader than ${\cal R}$.

\medskip
\noindent {\bf Example 3.1.} Let ${\cal A}$ be the
(commutative) algebra of $C^{\infty }$-functions on a smooth
manifold $M$.  Then ${\rm Der}{\cal A}$ is the Lie  algebra  of
smooth  vector fields on $M$. In this case $C_{\cal A}$ is the
algebra  of  continuous {\it locally} constant functions on
$M$. The dimension of $C_{\cal A}$ is then the  number of connected
components of $M$. So, $C_{\cal A}={\cal R}$ only if $M$ is
connected.

\medskip
\noindent {\bf Vectors.} Consider all
such $u\in {\rm Der}{\cal A}$ for which $V^{c}$ serves as the set of
constants, denote it $V^{cc}$. Clearly $V\subseteq V^{cc}$, however
this inclusion  may be strict. In the sequel we shall consider
such  collections $V$  of vectors that are uniquely determined by
their set of constants:

$$ V = V^{cc} \eqno{(3.3)} $$

\noindent
Such  requirement  looks  reasonable since  in  this  case $V$   is
automatically the Lie subalgebra of ${\rm Der}{\cal A}$. We shall
essentially  use this condition in the sequel (section 6).

\medskip
It follows from (3.3) that we could  define  the
differential algebra as a pair $({\cal A},{\cal K})$ putting
$V={\cal K}^{c}$ precisely as it was proposed by Geroch (1972).

\medskip
\section{CONNECTION AND CURVATURE}

\medskip
\medskip
In this section we  generalize  connection  to  non-commutative
differential algebras and introduce torsion and curvature.

\medskip
\noindent {\bf Connection.} In classical differential geometry
connection  provides the means to form the derivative of a vector
(field)  along  another one. We shall define it as a $V$-valued
function $\nabla _{x}y$ of two  arguments $x,y\in V$ such that it
is:

\medskip
\begin{enumerate}
\item {$S$-linear by the lower argument:
$\nabla _{zx}y = z\nabla _{x}y$ }

\item {$C_{\cal A}$-linear by the upper
argument: $\nabla _{x}(cy) = c\nabla _{x}y$ }

\item {Derivative with respect to $x$:
$$
\nabla _{x}(zy) = x(z)\cdot y+z\nabla _{x}y \eqno{(4.1)}
$$
}
\end{enumerate}

\noindent (recall that $S$, $C_{\cal A}$ are
the sets of  scalars  and  basic  constants, respectively).

\medskip
\noindent {\bf Torsion and curvature}. As in the classical
situation, the {\it torsion} is defined as the $V$-valued function

$$
T(x,y) = \nabla _{x}y-\nabla _{y}x-[x,y]\qquad x,y\in V
\eqno{(4.2)}
$$

\noindent It can be checked directly that $T$ is $S$-bilinear.  The
{\it curvature}  is defined as follows:

$$
R(u,x)y =\nabla _{u}\nabla _{x}y-\nabla _{x}\nabla _{u}y-\nabla _{[u,x]}y
\eqno{(4.3)}
$$

\noindent $R$ is the $V$-valued function of three arguments
$x,y,u\in V$. It  can  be  also verified that it is $S$-trilinear.
\par \medskip \noindent {\bf Ricci curvature.} In classical
geometry the Ricci curvature is formed as a contraction of the
curvature (4.3). Consider it in more detail.  Fixing up the values
$x,y$ in (4.3), we obtain the family of $S$-linear operators ${\cal
R}_{xy}:V\rightarrow V$

$$  {\cal R}_{xy}u = R(u,x)y \eqno{(4.4)}  $$

\noindent In the case when the notion of trace is meaningful for
operators  in $V$, the {\it Ricci curvature} is defined as the
trace of each operator ${\cal R}_{xy}$.

$$  {\rm Ric}(x,y) = {\rm Tr}{\cal R}_{xy} \eqno{(4.5)}  $$

\noindent {\bf The trace problem} is to define the trace in the
general situation as an $S$-linear  scalar-valued  functional  on
some  class  of  linear operators in $V$ such that

$$ {\rm Tr}(AB) = {\rm Tr}(BA) $$

When $V$ possesses a basis, the trace is defined as  the  trace  of
appropriate matrix. Whereas, even the module of vector fields $V$
may not have basis at all. For instance, in the  case  of
2-dimensional sphere ${\cal S}^2$, any smooth vector field on
${\cal S}^{2}$  has  at  least  one  point where it vanishes. Hence
no pair of vector fields can form the basis (but any three  vector
fields on ${\cal S}^{2}$  are  linear  dependent).  In classical
geometry we are always  in  a  position  to  localize  the
situation so that at {\it each point} the operator ${\cal R}_{xy}$
is  represented  by the matrix in a basis of the tangent space, so
that  the  trace  is well-defined.

\medskip
In the pointless situation all constructions are global. So, to
solve the trace problem along these lines, the basis is  to  be
set up. There are two obstacles, functional and algebraic, to do
it. The former is that the module $V$ may contain infinitely many
independent elements, the latter is that even finitely generated
module may  not have  basis.  A  possible  way  to  avoid  these
problem   is   the implementation of  algebraic  localization
(Atiyah  and  Macdonald, 1969).

\medskip
Another  approach  to  the  trace  problem   is   inspired   by
conventional Riemannian geometry.  It  is  based  on  the  canonical
isomorphism:

$$ V\otimes V^{*} \cong {\cal L}(V) \eqno{(4.6)} $$

\noindent where $V^{*}$ is the space of covector fields, and ${\cal
L}(V)$ is the  space  of linear (w.r.t. scalars) operators in $V$.
A  linear  operator  (4.4) represented as the element of $V\otimes
V^{*}$ is decomposed  into  the  sum  of the terms of the form
$v\otimes \omega$. The trace of each  term  is $\omega (v)$, hence
the trace of the entire operator is the sum of  appropriate values:
for $T\in {\cal L}(V)$

$$
T=\sum v_{i}\otimes\omega_{i}\hbox{  and }{\rm Tr}(T) = \sum \omega
_{i}(v_{i})
\eqno{(4.7)}
$$

\noindent In the general situation a more thoroughful  treatise  of  the
dualspace is needed.

\medskip
\section{COVECTORS}

\medskip
\medskip
Consider in more detail the algebraic structure of the  set  of
vectors $V$. First of all, $V$ is the real vector space.  Besides
that, it is equipped with the structure of two-sided $S$-module,
where $S$  is the set of scalars, which is assumed to be the center
$Z({\cal A})$  of  the basic algebra ${\cal A}$: for each $s\in S$,
$a\in {\cal A}$, $v\in V$

$$
a = s(va) = (v\cdot s)a = (va)\cdot s
$$

\noindent Note that the condition (3.3) is essential: otherwise $V$  would  not
possess the $S$-module structure.

\medskip
\noindent {\bf Covectors.} Now introduce the set of {\it covectors}
$V^{+}$ as

$$
V^{+}= {\rm Hom}(V,{\cal A})
$$

\noindent the set of all $S$-homomorphisms from the $S$-module $V$
to ${\cal A}$  considered $S$-module. $V^{+}$ is also the real
vector  space  and  possesses  the natural structure of ${\cal
A}$-bimodule: for each $\omega \in V^{+}$, $v\in V$, $a\in {\cal
A}$

$$
(\omega \cdot a)(v) = \omega (v)a
$$
$$
(a\cdot \omega )(v) = a\omega (v),
$$

The discrepancy between bimodule and two-sided module  is  that the
for  arbitrary $a\in {\cal A}$, $\omega \in V^{+}$ $\omega a\neq
a\omega $.  However  in  bimodules  the following holds:

$$
(a\omega )b = a(\omega b) \qquad
\hbox{,} a,b\in {\cal A} \hbox{,} \omega \in V^{+}
$$

\noindent {\bf Cartan differentials.} With each element $a\in {\cal
A}$ the  covector $da\in V^{+}$  is canonically associated:

$$
da(v) = v(a)
\eqno{(5.1)}
$$

\noindent The operator $d$ acts from ${\cal A}$ to $V^{+}$ (both
considered ${\cal A}$-bimodules)  so that the Leibniz rule holds:

$$
d(ab) = da\cdot b+a\cdot db
\eqno{(5.2)}
$$

It is the set of constants $V^{c}$ which is the kernel of  the
operator $d$. From the other side, not any covector may be of
the form $da$ for some $a\in {\cal A}$. For each $\omega \in V^{+}$
define its {\it Cartan differential} $d\omega $  as the following
skew symmetrical bilinear form on $V$:

$$
(d\omega )(v_{1},v_{2}) = v_{1}\omega (v_{2})-v_{2}\omega
(v_{1})-\omega ([v_{1},v_{2}])\qquad v_{1},v_{2}\in V
$$

\noindent When $\omega =da$ for some $a\in {\cal A}$, $d\omega $ is
necessarily  equal  to  zero.  However this is not the sufficient
condition.

\medskip
\noindent {\bf De Rham cohomologies.} A differential form $\omega $
is called {\it exact} if $\omega =da$ for some $a\in {\cal A}$, and
{\it closed} if $d\omega =0$. Since $dda=0$, each  exact  form
is closed. Both exact and closed forms are the submodules of
$V^{+}$,  hence their quotient can be formed called the module of
one-dimensional {\it De Rham cohomologies}
${\cal H}^{1}_{\cal A}(V)$.  In the classical case it  depends  on  the
topology of the underlying manifold ({\it e.g.} it  is  zero
for  simply connected manifolds). In  our  theory,  it  remains
the structural characteristic of the differential algebra $({\cal
A},V)$.

\medskip
The 0-dimensional cohomologies are defined as the algebra $V^{c}$
of constants. The closeness condition is now referred to the
elements of ${\cal A}: da=0.$ In virtue of (5.1) that means
that $va=0$ for each $v\in V$. In  the classical case ${\cal
H}^{0}_{\cal A}(V)$ is the number of connected components  of  the
manifold $M$ (see Example 3.1).

\medskip
\noindent {\bf Coupling.} There is the canonical coupling between
$V$ and $V^{+}$:

$$
<v,\omega > = \omega (v),\qquad v\in V, \omega \in V^{+}
\eqno{(5.3)}
$$

\noindent Due to non-commutativity we have  to  take  care  of  the
order  of factors:

$$
<v,a\omega > = a<v,\omega >; <v,\omega a> = <v,\omega >a \qquad
a\in {\cal A} \hbox{,} v\in V \hbox{,} \omega \in V^{+}
$$

\noindent The form $<*,*>$ is $S$-linear by the first and ${\cal
A}$-linear by  the  second argument. Thus any $v\in V$ can be
considered as ${\cal A}$-linear form on $V^{+}$:

$$ v\mapsto <v,*> \eqno{(5.4)} $$

In the classical theory all ${\cal A}$-linear  functionals  on
$V^{+}$  are exhausted by that of the form (5.4), which  does  not
hold  in  the general case. We shall deal with the class of  {\bf
regular  differential algebras} for which this also holds.

\medskip
\noindent {\bf Scalar covectors.} As it was already mentioned,  the
set $V$  is  the $S$-module. Its dual module is the set $V^{*}$ of
all $S$-{\it valued forms} on  $V$.  It is naturally to call
the elements of $V^{*}$ {\it scalar covectors}. Clearly
$V^{*}\subseteq V^{+}$, and moreover, $V^{*}$ is the $S$-submodule
of $V^{+}$.

\medskip
The canonical coupling (5.3) between $V$ and $V^{*}$ makes it
possible to consider the elements of $V$ as $S$-linear forms (5.4)
on $V^{*}$.  Note that each ${\cal A}$-linear form is $S$-linear,
but not vice versa,  hence  the regularity requirement does not
ensure us that any $S$-linear form  on $V^{*}$ is induced by some
$v\in ${\it V}. The module $V$ is called {\it reflexive} whenever
$V=V^{**}$.

\medskip
\noindent {\bf The trace problem again.} Return to the problem  of
representability of linear operators on by sums of terms $\omega
(v)$. We can not expect that the formula (4.6) will hold for
$V^{+}$ since $V^{+}$  is  broader  than  the "real" dual $V^{*}$.
Since $V^{*}$ is the submodule  of $V^{+}, (4.6)$  turns  to {\it
embedding}:

$$
{\cal L}(V) \cong  V\otimes V^{*} \subseteq  V\otimes V^{+}
\eqno{(5.5)}
$$

\noindent whenever $V$ is reflexive: $V=V^{**}$. In  this  case
the  trace  is  also well-defined in accordance with (4.7).

\medskip
Although, the regularity of $({\cal A},V)$ does not imply the
reflexivity of $V$. When $V$ is not reflexive, the elements
of ${\cal L}(V)$ are {\it approximated} by the elements of the
tensor product $V\otimes V^{*}$. To solve  this  problem, some
additional structure on $V$ or ${\cal A}$ must be imposed such
as  norm or topology. Although, the trace remains  non-uniquely
defined:  at least, up to a constant factor. This non-uniqueness
may change the form of the Einstein equation even in the classical
situation (cf.  section 7).

\medskip
\section{METRIC STRUCTURE}

\medskip
\medskip
The metric structure is introduced by  defining  an $S$-bilinear
${\cal A}$-valued symmetric form $g(u,v)$ on  $V$.  It
immediately  induces  the operator $V\rightarrow V^{+}$ defined for
any $v\in V$ as $v\mapsto \underline{v}$ such that:

$$
{v}(u) = g(u,v)
$$

We shall require the nondegeneracy of $g$, hence the mapping
$v\mapsto \underline{v}$ will be the injection. In the classical
situation it is isomorphism.  For general differential algebras
$v\mapsto \underline{v}$ is a mere embedding.

\medskip
\noindent {\bf Gradients.}  For  further  purposes  (to  introduce
the  Levi-Civita connection) we shall use weaker constraint than
the  requirement  of isomorphism $V\cong V^{+}$. Namely:

$$
\forall a\in {\cal A}\quad \exists v\in V \quad \underline{v}=da
\eqno{(6.1)}
$$

\noindent We shall call this vector $\underline{v}$ the {\it
gradient}  of the  element $a\in {\cal A}$  and denote it

$$
\underline{v} = {\rm grad}a \quad \hbox{ iff }\quad \underline{v}=da
$$

\medskip
\noindent The notion of gradient is unambiguously defined  in
virtue  of  the non-degeneracy of the metric form $g$.

\medskip
\noindent {\bf Levi-Civita  connection.}  In  the  standard
version   of   general relativity the Levi-Civita connection is
used.  That  is,  when  the metric $g$ is set up, the two following
conditions for the connection $\nabla $ hold for all $u,x,y\in V$:

$$
\nabla _{x}y-\nabla _{y}x = [x,y]
\eqno{(6.2)}
$$

$$
u(g(v,x)) = g(\nabla _{u}v,x)+g(v,\nabla _{u}x)
\eqno{(6.3)}
$$

\noindent The condition (6.2) means that $\nabla $ is
torsion-free: $T=0$,  and  (6.3) means that the covariant
derivative of $g$ is zero.

\medskip
In classical geometry the Levi-Civita  connection  is  uniquely
defined by the metric and always exists. Returning  to  the
general situation, let us try to build the connection  associated
with  the metric $g$.  First  suppose  it  exists.  Recall
how the  values  of Christoffel  symbols  are  obtained  in
classical geometry.   The variables $u$,$v$,$x$ are cyclically
permuted in (6.3) which yields  using (6.2):

$$
2g(x,\nabla _{v}u) = u(g(v,x)) + x(g(u,v)) - v(g(x,u)) -
$$
$$
\eqno{(6.4)}
$$
$$
 ( g(u,[v,x]) + g(x,[u,v]) - g(v,[x,u]))
$$

\medskip
\noindent To prove the existence, denote by $\Gamma (u,v,x)$ the
right side of (6.4).  Then fix up $u,v\in V$ and consider the
function $D_{v}u:{\cal A}\rightarrow {\cal A}$ defined as:

$$
D_{v}u(a) = \Gamma (u,v,{\rm grad}a)
$$

It can be checked directly that $D_{v}u$  enjoys  the  Leibniz rule
(4.1) and annihilates each  constant  from $V^{c}$,  hence the
mapping $a\mapsto D_{v}u(a)$  is  really  the  element  of $V$.
Moreover,  the  mapping $(v,u)\mapsto D_{v}u$ satisfies the
definition of connection (Section 4).

\medskip
Now we see the role of the conditions  (3.3)  and  (6.1):  they
enable the validity of the existence  theorem  for  the
Levi-Civita connection in differential algebras. The uniqueness
follows from the non-degeneracy of $g$.

\medskip
\section{EINSTEIN EQUATION}

\medskip
\medskip
Now we have everything to introduce the point-free  counterpart
of the Einstein equation . In conventional theory it postulating the
equality between the Einstein tensor depending on geometry only  and
the momentum-energy tensor.

\medskip
To form the left side, first introduce the analog of the scalar
curvature $R$. In classical geometry $R$  is  the contraction of
the contravariant metric tensor with the Ricci tensor.  In
differential algebras  we  have  neither  contraction nor tensors,
but   only operators. Although we have the trace of operators in
our  disposal.  It worked already when the Ricci operator was
defined as  the  trace of Riemann curvature (4.5).

\medskip
\noindent {\bf The Ricci operator and scalar curvature.} In (4.5)
the $S$-bilinear form ${\rm Ric}(x,y)$ was defined. To define the
scalar curvature we  must be in a position to associate the form
{\rm Ric} with an operator ${\cal R}$ in $V$ such that:

$$
{\rm Ric}(u,v) = g({\cal R}u,v)
\eqno{(7.1)}
$$

\noindent Then the trace of ${\cal R}$ will be the scalar curvature
$r$

$$
r = {\rm Tr}{\cal R}
\eqno{(7.2)}
$$

Whereas (7.2) is well defined  only  of  (i)  this  operator ${\cal
R}$ exists and (ii) ${\cal R}$ will be trace-class operator w.r.t.
the trace  ${\rm Tr}$.  Leaving apart the item (ii), we suggest a
sufficient condition for ${\cal R}$ to exist. For each $u\in V$
define the covector $u_{\cal R}$ as:

$$
u_{\cal R}(v) ={\rm Ric}(u,v)
\eqno{(7.3)}
$$

\noindent Now if we require for any $u\in V$ the existence of $v\in
V$ such that:

$$
u_{\cal R} = \underline{v}
\eqno{(7.4)}
$$

\noindent the operator ${\cal R}$ can be immediately defined as
${\cal R}u=u_{\cal R}$.

\medskip
\noindent {\bf The Einstein equation.} In conventional relativity
the operator  form of the Einstein equation is:

$$
R^{i}_{k}- {1\over 2}R\delta ^{i}_{k} = \kappa T^{i}_{k}
\eqno{(7.5)}
$$

In differential algebras $R^{i}_{k}$ turns to the Ricci operator
${\cal R}$  and the scalar curvature is $r$. So, everything
is now ready to write  the analog of (7.5):

$$
{\cal R} - {1\over 2}rI = \kappa {\cal T}
\eqno{(7.6)}
$$

Note that (7.6) substantially depends  on  the  choice  of  the
trace, however so does (7.5)! It is assumed in  the  classical
case that the trace of the unit operator $\delta ^{i}_{k}$ is equal
to 4 (the  dimension of spacetime manifold). Although, we could
redefine  the  trace  so that all contractions would be multiplied
by a constant $\alpha $,  and  the factor 1/2 in (7.5) will turn to
$1/2\alpha $.

\medskip
We reproduce the Einstein equation (7.5)  in  the  form  (7.6),
which requires the introduction of the momentum  operator ${\cal
T}$,  which acts as follows. Recall that $V$ is interpreted as the
set of  virtual shifts, So, if $v\in V$ is associated with a
shift of the  observer, ${\cal T}v$ yields the energy flow he
observes.

\medskip
\section{AN EXAMPLE}

\medskip
\medskip
Consider the basic algebra ${\cal A}={\rm Mat}_{4}({\cal R})$ of
square $4\times 4$  matrices, and let ${\cal K}$ be the subalgebra
of ${\cal A}$ generated by the matrices:

$$
{\bf e}_{14} =\left(\matrix{
0 & 0 & 0 & 1\cr
0 & 0 & 0 & 0\cr
0 & 0 & 0 & 0\cr
0 & 0 & 0 & 0}\right) \qquad
{\bf e}_{24} =\left(\matrix{
0 & 0 & 0 & 0\cr
0 & 0 & 0 & 1\cr
0 & 0 & 0 & 0\cr
0 & 0 & 0 & 0}\right)
\eqno{(8.1)}
$$

Denote by ${\bf e}_{ik}$ the matrix having 1 at the $(i,k)$ entry
with  all other entries equal to zero. In accordance with (3.3),
form the  set $V={\cal K}^{c}$. Any derivative in ${\rm
Mat}_{n}({\cal A})$ is inner, that is, each  element  of $v\in V$
is associated with a matrix $v\in {\cal A}$ so that $v(a)\equiv
va-av=[v,a]$ for any $a\in {\cal A}$. The Lie  operation  in $V$
is  the  commutator  of  appropriate associated matrices.

\medskip
\noindent It can be checked by direct calculation that $V$ is the
6-dimensional Lie algebra spanned on the elements $\{{\bf
e}_{13},{\bf e}_{14},{\bf e}_{23},{\bf e}_{24},{\bf
e}_{33},{\bf e}_{34}\}$.  The nonzero commutators of the basis
elements are:

$$
[{\bf e}_{13},{\bf e}_{33}]={\bf e}_{13}\hbox{;}\quad [{\bf
e}_{13},{\bf e}_{34}]={\bf e}_{14};
$$

$$
[{\bf e}_{23},{\bf e}_{33}]={\bf e}_{23}\hbox{;}\quad [{\bf
e}_{23},{\bf e}_{34}]={\bf e}_{24}\hbox{;}\quad [{\bf e}_{33},{\bf
e}_{34}]={\bf e}_{34}
$$

\noindent and other commutators are zero. The set of scalars $S$
of  the  model will be the set of scalar matrices $\lambda I$,
where $I$ is  the  unit  matrix and $\lambda \in {\cal R}$.

\medskip
Suppose  that  some  metric  structure  on $({\cal A},V)$  is
defined:  $g:V\times V\rightarrow {\cal A}$. In accordance with
(7.1) the  values  of $g$  must  be  the values of the trace, {\it
i.e.} scalars. Hence $g(u,v)=\lambda I$.

\medskip
Now try to build the Levi-Civita connection associated with the
metric $g$. To do it, we must check the existence of gradients
(6.1).  That means that for all $a\in {\cal A}$ such $v\in V$ must
exist that for all $u\in V$

$$
g(v,u) = da(u) = u(a) = [u,a]
$$

\noindent Hence the commutator $[u,a]$ is a multiple of the unit
matrix. This is possible in {\it no finite dimensional} case.
Although, it may be possible in infinite-dimensional space where
canonically conjugated variables do  exist.  Nevertheless,
torsion-free  connections  exist  in  our example, for instance,
that defined as:

$$
\nabla _{u}v = uv
$$

\noindent It can be checked directly that the conditions (4.1) are
valid  due to that the scalars are multiples of unit matrix.

\medskip
This example shows that the {\it affine} differential structures  can
survive  even  on  finite-dimensional  basic  algebras,  while   the
attempts to build the non-commutative  {\it Riemannian}  geometry  require
the  infinite  dimensionality  of  basic  algebras.  Moreover,   any
attempts to substitute spacetime by finitary  patterns  can  restore
either metric (Regge calculus) or topology (Zapatrin, 1993), but not
both at once.

\medskip
\noindent {\bf SUMMARY AND CONCLUDING REMARKS}

\medskip
\medskip
\noindent Begin with an outline of main results of our paper:

The models of point-free differential geometries  are  proposed
(section 3) as pairs $({\cal A},V)$, called {\it differential
algebras}  which  are the non-commutative generalization  of
Einstein  algebras  (Geroch, 1972).  The  substantial  feature  of
non-commutativity   is   the discrepancy between the elements of
the basic algebra (analog of the smooth functions) and scalars
(section 2).

\medskip
Luckily,  the  geometry  of  affine  connection   survives   in
non-commutative differential  algebras,  including  the  notions  of
torsion and curvature (section 4). It is even possible to  introduce
"topological" invariants such as De Rham cohomologies  (section  5).
The conditions for the Ricci form to exist was reduced to the  trace
problem. The possible ways to solve it are shown.

\medskip
The conditions for a  metric  structure  to  be  definable  are
studied giving rise to the notion of {\it regular}  differential  algebras
(section 6). It happens so that the  Levi-Civita  connection  (which
can always be restored from  the  symmetrical  metric  form  in  the
classical case) may not exist in non-commutative  case  (example  in
section 8):  it  depends  of  the  possibility  to  build  gradients
(section 6). Whereas, if it exists, it is still unique.

\medskip
The  scalar  curvature  can  also  be  defined  under   certain
circumstances. If  it  becomes  possible,  the  operator  analog  of
Einstein equation is introduced. It is shown that it does not depend
on the normalization of the trace.

\medskip
The idea to consider vectors as differential operators  applied
to functional algebras, but defined on a  broader  class  of
spaces than manifolds, called {\it differential spaces}, were used
to  implement the spaces with singularities to general relativity.
Heller  {\it et  al.} (1989) have shown that the  reasonable
definition  of  differential structure can be  formulated  in
terms  of  certain  algebra ${\cal A}$  of functions so that even
the analog of Lorentz structure can be introduced (Multarzinski
{\it et al}, 1990).  In particular, when ${\cal A}$ is an algebra
of smooth functions on a manifold, the standard differential
geometry is restored.

\medskip
The approach we suggest can be considered as a  reasonable  way to
quantize the gravity. The main problem arising here  is  to  find
the appropriate representations of the basic  algebras.  What
about the source of basic algebras, this is the  Wheeler's
suggestion  to consider  logic  as  pregeometry  which  could  work
here.  The first  step along these lines was made by  Isham
(1989):  the  lattice  of  all topologies over a set was
considered, and the analog of creation and annihilation operators
was suggested. The appropriate algebra  could be taken as basic
one. Moreover, starting from an arbitrary property lattice as
background object, one can  always  build  the  semigroup (called
{\it generating},  Zapatrin,  1993b)  whose  annihilator lattice
restore this property lattice. Then  the algebra spanned on this
semigroup could play the role of  the  basic algebra ${\cal A}$.

\medskip
The next step is the  {\it spatialization}  procedure  (mentioned
in section 2). When a differential structure $V$ and a metric $g$
are  set up, the problem arises to extract usual (i.e. point)
geometry  from the triple $({\cal A},V,g)$.  To  return  to
points,  we  must  consider  a subalgebra ${\cal C}\subseteq {\cal
A}$  such  that ${\cal C}$  would  be  commutative  (to   enable
functional representation) and in some sense concerted with $V$ and
$g$.  We did not yet tackle this problem of {\it eigen-subalgebras}
in  detail, whereas it looks as direct way to reveal events within
our  scheme.  It is noteworthy that whenever the triple $({\cal
A},V,g)$ is set  up,  there still    may exist     several
functionally     representable eigen-subalgebras   associated
with    possibly    non-isomorphic geometries.  That  means  that
{\it the  observed  geometry  depends  on observation} which is in
complete accordance with quantum  mechanical point of view.

\medskip
\noindent {\bf ACKNOWLEDGMENTS}

\medskip
\medskip
We appreciate the attention to our work  offered  by  A.A.Grib.
One of the authors (R.R.Z.) takes the opportunity to  acknowledge  the
profound discussions with D.Finkelstein (on what  quantum  spacetime
ought to be) and M.Heller (on Einstein  algebras)  and  a  financial
support  from  ISF  (George  Soros  Emergency  Grant)   and   Pavlov
Enterprise (St-Petersburg).

\medskip
\noindent {\bf REFERENCES}

\medskip
\noindent Atiyah, M.F. and Macdonald, I.G. (1969),
{\it Introduction to commutative algebra},
Addison-Wesley, Reading, Massachusetts
\medskip

\noindent Geroch, R. (1972),
Einstein Algebras,
{\it Communications in Mathematical Physics},
{\bf 26},
271
\medskip

\noindent Hawking, S.W., and Ellis G.F.R.(1973),
{\it The Large-Scale Structure  of Space-Time},
Cambridge University Press, Cambridge
\medskip

\noindent Heller, M., Multarzinski, P., and Sasin, W.  (1989),
The  algebraic approach to space-time geometry,
{\it Acta Cosmologica},
{\bf 16},
53
\medskip

\noindent Isham C.J.(1989),
Topology lattice and the quantization on the lattice of topologies,
{\it Classical and Quantum Gravity},
{\bf 6},
1509
\medskip

\noindent Multarzinski, P., and Heller, M. (1990),
The Differential and Cone Structure of Spacetime,
{\it Foundations of Physics},
{\bf 20},
1005
\medskip

\noindent Zapatrin, R.R. (1993a),
Pre-Regge  Calculus:  Topology  Via  Logic,
{\it International Journal of Theoretical Physics},
{\bf 32},
779
\medskip

\noindent Zapatrin, R.R. (1993b),
Generating Semigroups for Complete Atomistic Ortholattices,
{\it Semigroup Forum},
{\bf 47},
36

\end{document}